\documentclass[  a4paper,
                 ,jap
                 ,twocolumn
                 ,superscriptaddress
               ]{revtex4}
\usepackage{amsmath}
\usepackage{amsfonts}
\usepackage{amssymb}
\usepackage[english]{babel}
\usepackage{fontenc}
\usepackage{graphicx}
\usepackage{natbib}
\usepackage{hyperref}

\hypersetup{colorlinks=true,linkcolor=blue,citecolor=blue,pdftitle={Numerical and experimental studies of the carbon etching in EUV-induced plasma}}

\newcommand*{\doi}[1]{\href{http://dx.doi.org/\detokenize{#1}}{doi:\detokenize {#1}}}

\graphicspath{{images/}}

\begin{document}
\title{Numerical and experimental studies of the carbon etching in EUV-induced plasma}
\author{D.I.~Astakhov}
\affiliation{XUV Group, MESA+ Institute for Nanotechnology, University of Twente, P.O. Box 217, 7500 AE Enschede, The Netherlands}
\author{W.J.~Goedheer}
\affiliation{FOM Institute DIFFER - Dutch Institute for Fundamental Energy Research, PO Box 6336, 5600HH Eindhoven, the Netherlands, The Netherlands}
 \author{C.J.~Lee}
\author{C.J.~Lee}
\affiliation{XUV Group, MESA+ Institute for Nanotechnology, University of Twente, P.O. Box 217, 7500 AE Enschede, The Netherlands}
\author{V.V.~Ivanov}
\author{V.M.~Krivtsun}
\author{O.~Yakushev}
\author{K.N.~Koshelev}
\affiliation{Institute for Spectroscopy RAS (ISAN), Fizicheskaya 5, Troitsk 142190, Russian Federation}
\author{D.V.~Lopaev}
\affiliation{Skobeltsyn Institute of Nuclear Physics, Lomonosov Moscow State University, Leninskie Gory, Moscow 119991, Russian Federation}
\author{F.~Bijkerk}
\affiliation{XUV Group, MESA+ Institute for Nanotechnology, University of Twente, P.O. Box 217, 7500 AE Enschede, The Netherlands}

\begin{abstract}
We have used a combination of numerical modeling and experiments to study carbon etching in the presence of a hydrogen plasma.  We 
model the evolution of a low density EUV-induced plasma during and after the EUV pulse to obtain the
energy resolved ion fluxes from the plasma to the surface. By relating the
computed ion fluxes to the experimentally observed etching rate at various pressures and ion energies, we show that at low pressure and energy, carbon etching is due to chemical sputtering, while at high pressure and energy a reactive ion etching process is likely to dominate. 
\end{abstract}
\maketitle
\section{Introduction}
Many vacuum mirror optical tools suffer from the build-up of carbon
contamination due to cracking of hydrocarbons under powerful vacuum
ultraviolet  radiation~\cite{Boller.1983.investigation}. The large absorption of EUV radiation by carbon becomes significant in the case of multi-element optical systems, where throughput can be greatly reduced by even a very thin layer of carbon contamination on the top of each mirror. 

The problem of EUV induced carbon contamination has been addressed in a series of publications \cite{Hollenshead.2006.modelinga,Nakayama.2009.analysis,Shin.2009.reflectivity,Davis.2007.situ}. The reduction of carbon films in a hydrogen atmosphere or hydrogen plasma has also been extensively studied~\cite{Hopf.2003.chemical,Kuppers.1995.hydrogen,Liu.2010.general, Jariwala.2009.atomic,Dolgov.2013.comparison}. Despite numerous studies, however, it is still difficult to predict the carbon removal rate, because there are many contributing factors. Several aspects that significantly affect the carbon removal rate include: many different allotropes and compounds of the carbon (e.g. soft black or hard graphite),  many different contributing reaction paths (e.g. physical sputtering, chemical sputtering, reactive ion etching etc.), and, last but not least,  small admixtures to the background gas, which, while residing on the carbon surface, can produce reactive species once irradiated by EUV. Despite this complexity, experiments have shown that carbon etching can be achieved under certain EUV-induced plasma conditions. Nevertheless, it has proven difficult to fully understand the etch process, because the characteristics of the EUV-induced plasma are poorly known, and the plasma-surface interaction has many contributing factors~\cite{Dolgov.2013.comparison}. 

In this paper, we use a model of the EUV-induced plasma  to numerically analyse the fluxes from the plasma to the sample surface. Our model is a self-consistent 2D Particle-in-Cell model of the weakly ionized low pressure hydrogen plasma that is formed during the EUV pulse due to ionization by EUV photons and secondary electrons from the surface. As described in Section~\ref{section:model}, our model provides an accurate estimate  the ion flux composition and energy distribution.  However, a considerable number of parameters required for accurate simulations are not well known, therefore, the modeling results were combined with experimental observations. With  the aid of simulations, we  show that the shape of the energy distribution function of the ion fluxes in the considered experimental setup  are mainly defined by the setup geometry, background pressure, and externally applied bias voltage. The ion dose, on the other hand  is sensitive to the variations of many other parameters (e.g. EUV dose, secondary electron yield, etc.). 

By  combining  the computed energy distribution function of the ion flux with the experimentally measured ion dose,  insight into the  mechanism for carbon removal was gained. By analysing the differences in yield between EUV-induced plasma and surface wave discharge plasma experiments in combination with numerical simulations, we show that chemical sputtering dominates for low pressures and energies. It was found that the carbon removal yield for both the surface wave discharge and EUV-induced plasmas was similar in the overlapping energy range. Hence, the effect of the EUV radiation on carbon removal is found to be significantly smaller than was estimated previously~\cite{Dolgov.2013.comparison}.

\section{Experimental setup \label{section:experiment}}

\begin{figure}    
  \centering
  \includegraphics[width=0.8\columnwidth]{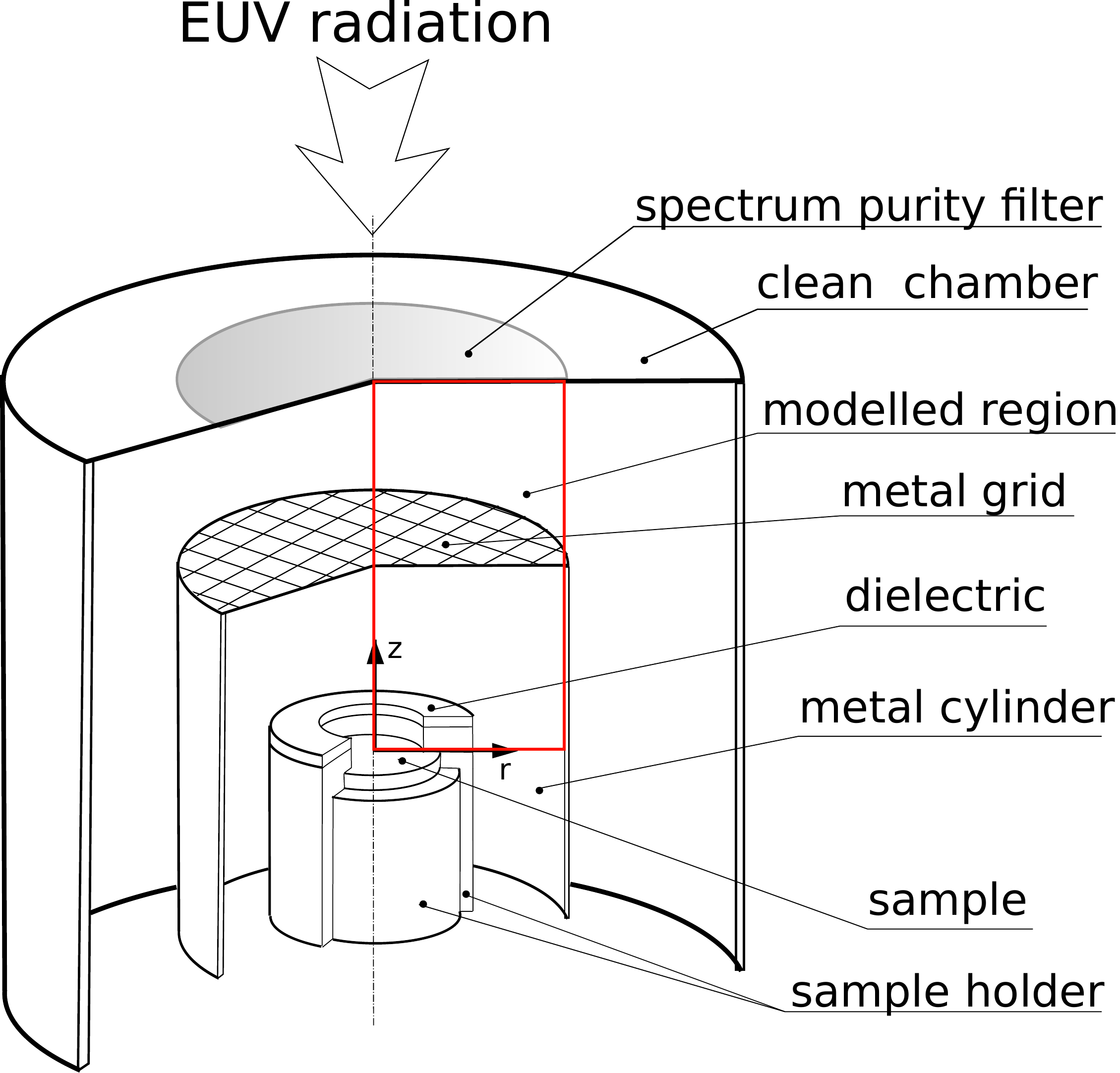}
  \caption{Configuration of the experimental chamber. The metal cylinder inner diameter is 28~mm, the sample holder diameter is 18~mm, the inner radius of dielectric mica diaphragm is 8~mm, the distance from sample to grid is 2.4~cm, distance between the grid and SPF is 1.5~cm.  \label{fig:proto_chamber}}        
\end{figure}

The ISAN EUV experimental setup is based on a tin EUV radiation source, which is a Z-pinch discharge plasma with $1500$~Hz repetition rate, which has been described in detail elsewhere~\cite{Dolgov.2015.extreme}. This source is a good tool for exploring EUV-induced surface processes over a large number of pulses ($>$~1~MShot).
In brief, EUV radiation is introduced into a so-called "clean" chamber (see Fig.~\ref{fig:proto_chamber}), separated from the source and collector optics by a Si:Zr spectral purity filter (SPF). The clean chamber is differentially pumped to pressure of \mbox{$3.5\times10^{-8}$~torr}. Under vacuum conditions, the background hydrocarbon carbon growth rate was measured to be 0.4$\pm$0.2~nm/10~Mshot.

The diameter of the EUV beam at the sample was 5~mm. In addition to the direct beam, some scattered EUV radiation was also incident on the sample. The EUV pulse duration is about 100~ns (FWHM), with a tail, as described in~\cite{Bakshi.2006.euv_sources}.

The incident EUV power was measured using a sensitive thermo-couple attached to a thin copper disk. It was found that the radiation intensity was  $0.13$~W/cm$^2$ after the SPF and approximately $0.75$~W/cm$^2$ without the SPF. The ratio of EUV intensities with and without the SPF corresponds to the calculated transmission of the  100~nm Si:Zr SPF filter over an energy range of 60 to 100~eV (see Fig~\ref{fig:euv_spectrum}).

The EUV intensity on the sample decreases with time because of  carbon growth on the SPF filter and focusing optics. In later experiments, the EUV intensity was measured to be 0.1~W/cm$^2$ after the SPF.   

During the experiments,  the hydrogen pressure was set in the range of 2.8~Pa -- 86.5~Pa. To ensure that each radiation pulse excited a plasma in an atmosphere dominated by hydrogen,  hydrogen flowed through a liquid nitrogen trap (to remove water) and into the chamber at 100~liters$\times$torr/minute .  

To control the energy of the ion flux,  the sample holder assembly was biased in the range of -200~--~0~V, while all other metallic electrodes were grounded. The samples consisted of carbon, deposited by magnetron sputtering on a silicon wafer to a thickness of $\sim$30~nm. Each sample was exposed to the  10$^7$ EUV pulses at a different combination of bias and hydrogen pressure. After exposure, the amount of carbon removed was measured by X-Ray fluorescence (XRF) (EDS) analysis and spectroscopic ellipsometry.

\section{Model \label{section:model}}
A two dimensional particle in cell (PIC) model with $rz$ geometry was used to
model the experiment. Our model follows the general PIC scheme, described elsewhere~\cite{Birdsall.1985.plasma}.

Ionization is initiated by an EUV pulse, which directly ionizes the background gas, and produces electrons by photo emission from the SPF and the sample. This process leads to the formation of an EUV induced plasma in the chamber. The ionization and photo-emission process and their inclusion in the model is described below.

Although the plasma is continually re-ignited by the pulsed EUV radiation, we consider each EUV pulse to induce a plasma in a cold neutral gas that is at equilibrium, because the  characteristic time of plasma decay in the given
geometry is about $\sim 20\mu s$, which is much shorter than the 660~$\mu$s between
pulses. The restoration of thermal equilibrium  in the background gas is also much shorter then the time between pulses. 

\subsection{Chamber configuration}
The configuration of vacuum chamber is presented in Fig.\ref{fig:proto_chamber}. Because the chamber is axially symmetric, it is possible to include the structure of the internal chamber in the simulations. To accurately model the plasma dynamics when the sample is biased, the space between the SPF and the metal grid is included in our simulations. The grid is included in the model as a number of metallic rings.  When a bias is applied to the sample, ions from the plasma, which is formed between SPF and metal grid, are pulled towards the sample. 

\subsection{Deielectric model}
In the experiment, a dielectric mica diaphragm was used to prevent EUV radiation being incident on the sample holder structure, thereby decreasing electron
emission from the sample holder structure. Therefore, allowing plasma parameters to be estimated from the discharge characteristics of the sample. In the model, this feature is included with a simple dielectric model: a dielectric is not conductive, it can accumulate charge and the secondary electron yield under EUV irradiation is order of magnitude lower than for the sample. 

The electron reflection and secondary electron emission under electron impact (SEE) are included in the model. These two processes are combined into SEE with the probability ($P(E)$) defined as follows.
\begin{equation}
    P(E) = p_0 + \frac{E_e}{E_1}(1.0 - p_0) 
\end{equation}
Here E -- is the energy of incoming electron. The parameters of the used mica are unknown, therefore, we choses $E_1 = 45$~eV and $p_0 = 0.7$, these parameters corresponds to linear fit for SEE yield for quarts in \cite{Dunaevsky.2003.secondary}. 

This approximation blends between significant slow electron backscattering for low energy electrons and secondary electron emission for higher impact energies. We do not use higher order approximations, since both effects are known to depend strongly on the surface conditions~\cite{Fowler.1958.reflection,Dunaevsky.2003.secondary}.

This model for the dielectric is over simplified, but sufficient to allow us to to take the out of focus EUV radiation, which  is incident on the diaphragm, into account in a consistent manner.

\subsection{Length scales and grid resolution \label{section:length_scales}}
The time evolution of one EUV pulse has two distinct stages: negative
space-charge dominated during the beginning of the EUV pulse, and decay of the positively charged plasma after the pulse \cite{vanderVelden.2006.particle-in-cell}.

In order to model the negative space-charge dominated part of the plasma evolution,  the potential well near the surface, which is formed by the electrons that have escaped to the volume, must be resolved. The length scale of the space-charge potential well depends on the energy distribution of photoelectrons from the surface and on the current which passes through the system.

For the purpose of estimation, it is possible to simplify the formulas from \cite{Langmuir.1923.effect} and obtain
\begin{equation}
   z_m \simeq 0.1\text{ cm} \times \frac{(T[\text{eV}])^{3/4}}{(I[\text{mA/cm$^2$}])^{1/2}}
\end{equation}
Here, $T$ is the initial temperature of the emitted photoelectrons, $I$ is the current density near the cathode, and $z_m$ is the distance from the cathode to the bottom of the space charge potential well. For $T\sim$1~eV and $I \sim$20~mA/cm$^2$, one obtains  $z_m \sim$0.02~cm. 

For the plasma dominated part of the plasma evolution we need to resolve the Debye length in the volume, and also have a finer grid near the surface to resolve the plasma sheath.

From the EUV intensity  ($\sim$0.02~mJ per pulse) and background hydrogen pressure
(2.8~Pa, 11.2~Pa),  the plasma density can be estimated to be about
10$^9$~cm$^{-3}$. For $T_e \sim$0.5~eV and $N_e \sim$ 10$^9$~1/cm$^3$ one obtains  $R_D \sim$0.015~cm, which is comparable to the space charge length scale estimated above. To achieve this, the grid is refined near the sample surface and SPF. The minimum grid cell size is $0.05\times R_D$, and is gradually increased in steps of 5\% until the bulk cell size of $0.5 \times R_D$ is reached. Several tests were performed to ensure that the chosen grid resolution does not affect the plasma dynamics.

\subsection{photo-electron emission \label{section:SEE}}
The energy spectrum of photoelectrons emitted from the SPF and sample surface are calculated from~\cite{Henke.1977.electron_emission}: 
\begin{equation}
P(E) \sim \frac{E}{(E + W)^4} 
\label{eqn:se_spectrum}   
\end{equation}
Here $P(E)$ is the probability of emitting an electron with energy, $E$, from a surface with a work function, $W$. For carbon  $W=5$~eV~\cite{Lide.2003.crc}. For electrons emitted from the surface, we assume an angular dependence given by a cosine emission law~\cite{Henke.1977.electron_emission}. For purpose of estimation, we used the same approximation for the spectrum of photoelectrons emitted from the dielectric.

\subsection{EUV spectrum and photoionization}
To calculate the distribution of ion species, the spectrum of the EUV radiation  must be included in the model, along with the energy-dependent cross sections for each photoionization process. Direct photoionisation of hydrogen by EUV photons is included as two  processes:
\begin{eqnarray*}
    \mbox{H}_2 + h \nu & \rightarrow & \mbox{H}_2^+ + e  \\
    \mbox{H}_2 + h \nu & \rightarrow & \mbox{H}^+ + \mbox{H} + e 
\end{eqnarray*}
The photoionisation cross-sections were taken from \cite{Chung.1993.dissociative}. 

\begin{figure}
  \includegraphics[width=\columnwidth]{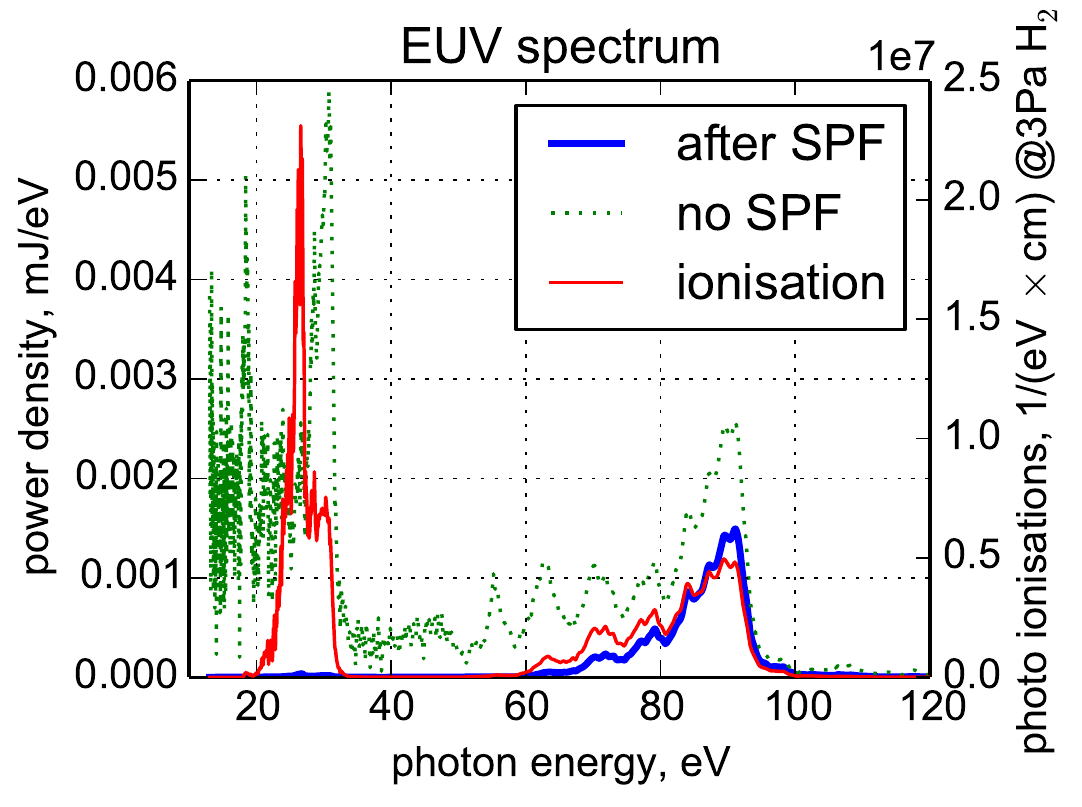}
  \caption{EUV spectrum after SPF as used in the simulations. Note, that the small SPF  transmission,  estimated at $\sim$1\% of EUV energy per pulse,  significantly contributes to the number of photoionisation events in the volume. \label{fig:euv_spectrum}}
\end{figure}

In the experiment, the EUV spectrum was measured (green curve Fig.~\ref{fig:euv_spectrum}) and found to be in agreement with data from~\cite{Bakshi.2006.euv_sources}. The EUV spectrum, however, was measured before reflection from  the Zr collector mirrors, and transmission through the SPF.

The calculated transparency curve that was provided with the Si:Zr SPF was used to calculate the EUV spectrum after transmission through the SPF (See Fig.~\ref{fig:euv_spectrum} blue curve).  

The SPF transmission curve has a transmission of about 1\% in the range of 20~--~30~eV, which is important because the photoabsorption cross-section is very large in this energy range (see Fig.~\ref{fig:euv_spectrum} red curve). 

The contribution due to radiation in the 20~--~30~eV range is difficult to quantify for two reasons: the intensity of the radiation varies significantly due to carbon growth on the SPF, and the accuracy of the SPF transmission curve in this energy range is unknown. Nevertheless, we include it because, even a small transmission will result in  additional photoinization, which leads to a reduced space-charge potential barrier at low pressures (e.g. 2.8~Pa, 11.2~Pa). 

Although the photoionization due to the 20~--~30~eV radiation range is significant compared to the 13~nm band, it has significance only for cases where no bias is applied to the sample. For cases with bias, the presence of this radiation leads to a decrease of the space-charge potential. 

\subsection{Cross-sections set}
The cross-sections set used in the  model consists of electron collisions with hydrogen and ion collisions with hydrogen. The collisions between plasma species and three body processes are neglected due to their low probability under the conditions considered here. To accurately model electron collision related processes in hydrogen discharges with a  Monte Carlo (MC) model, one needs to take into account the differential cross-sections for ionisation and excitations processes. As described in detail in \cite{Mokrov.2008.monte_carlo}, the particular choice of the angular dependence of cross-sections significantly influences the simulation results. We adopt a set of cross sections found in Ref.~\cite{Mokrov.2008.monte_carlo}.

We use an approach similar to that described in \cite{Mokrov.2008.monte_carlo}. The reaction probability is sampled using the integrated reaction cross-section. The differential cross-section data is used to determine the collision kinematics and energy redistribution between products. Electron elastic scattering and hydrogen electronic excitations, and angular scattering data is taken from \cite{Brunger.2002.electronmolecule}. For electron impact ionization of hydrogen, we use an experimentally determined doubly differential cross-section \cite{Shyn.1981.doubly,Rudd.1993.doubly}. The set of cross-sections for collisions between ions and hydrogen is based on~\cite{Simko.1997.transport} because this set provides good agreement with swarm data for ions in hydrogen. We neglect the formation of H$^-$, because the cross-section of dissociative electron attachment is very low, and the density of vibrationally exited hydrogen molecules too low to make a significant contribution to the production of H$^-$.

We make use of the procedure described in \cite{Nanbu.1994.simple} to perform Monte Carlo
collisions with the background gas. We tested the consistency of our implementation by modeling swarm experiments and found good agreement with experimental values~\cite{Dutton.1975.survey} for the first Townsend electron ionization coefficient, the electron mobility, and for H$^+$  and  H$_3^+$ mobility in hydrogen \cite{Graham.1973.mobilities}.
 
\section{Analysis of the charge --- bias characteristic\label{section:qv_characteristics}}
\begin{figure}
  \centering
  \includegraphics[width=\columnwidth]{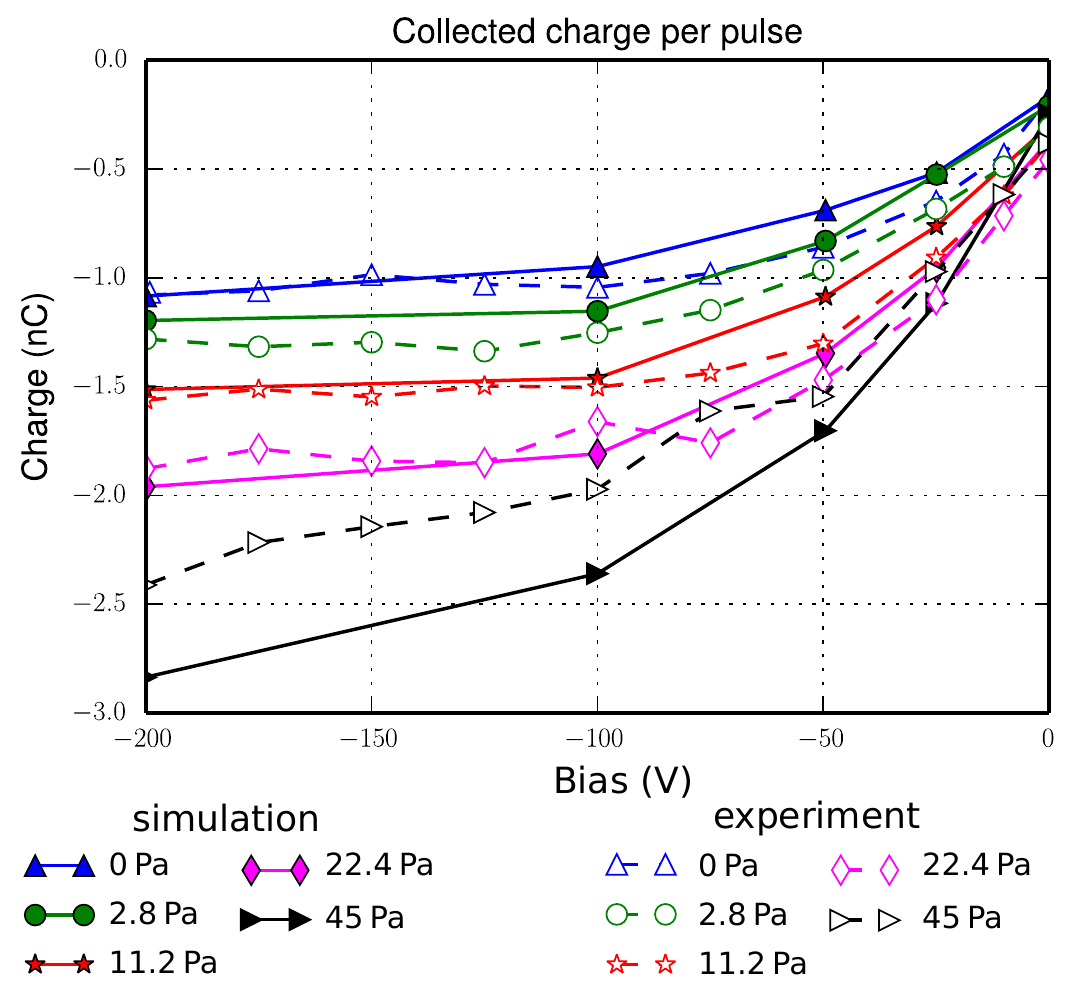}
  \caption{Collected charge as function of bias, $0.017$mJ/pulse ($0.13$ W/cm$^2$  at 1500~Hz to the spot of 5~mm diameter. \label{fig:IV}}
\end{figure}

Let us begin with the analysis of the charge~---~bias voltage characteristics (Q~--~V), which were measured during the experiments. The ion dose on the sample per pulse for the range of pressures and bias voltages considered here
was estimated from the experimental results as a difference between the collected charge for given pressure and bias voltage and the collected charge for vacuum conditions \cite{Dolgov.2013.comparison}. The measured curves are presented in Fig.\ref{fig:IV}.   It is worth noting that these $Q-V$ characteristics have some peculiarities. 

Firstly, all the characteristics reach saturation, and the turning point for the almost all curves is approximately -75~V. If there is a significant contribution from  ionization by the accelerated secondary electrons from the sample in the drift regime, one would expect a rapid increase of the collected charge for biases in the range of -200~..~-50~V.

Secondly, the increase of the collected charge for -200~V bias does not depend linearly on pressure in the range of 2.8~Pa to 22.4~Pa, i.e. the collected charge increases less than two-fold while the pressure increases eight times. But, one would expect at least linear growth of the collected charge with  pressure if the volume ionization provides a significant contribution to the total collected charge. Thus, the overall contribution of photoionization in the volume is small  compared with the other factors.

Thirdly, for -200~V bias and 22.4~Pa background pressure, due to the photoionization process, all the external bias would be applied over a  very small layer near the sample, thus, all the secondary electrons will gain 200~eV energy. The H$_2$ ionization cross-section for this energy is approximately $6\cdot 10^{-17}$~cm$^2$, thus, in the space between the sample and the grid ($l\sim 2.5$~cm), these electrons will, on average, have 0.8 ionization inducing collisions. If we assume that, under vacuum conditions and -200~V bias, the SE current from the sample saturates, and we obtain that, just due to the direct electron induced ionization, without cascade process and photoionization, the collected charge should be greater than 1.8~nC. in the experiment, however, the collected charge is approximately 1.9~nC. 

\subsection{Average secondary electron yield }
The above analysis suggests that the combination of EUV power per pulse, and effective secondary electron yield (SEY) produces approximately 1~nC of electrons from the sample. 

It is instructive to estimate the total SE charge from the experimental parameters. From the EUV intensity, repetition rate and spot size, the average dose per pulse was approximately $\sim$0.017~mJ. The secondary electron yield for carbon under EUV radiation is estimated to be approximately \mbox{$\gamma_{SE} \sim 0.01$} for the photon energy range of 60~eV~--~100~eV~\cite{Yakshinskiy.2007.carbon, Hollenshead.2006.modelinga}, which leads to a saturation charge of 2~nC, which would lead to a significant disagreement between the simulated and measured charge bias characteristics. It is also worth noting that effective SEY from mica was small, since, in the experiment, the charge collected from a sample made from mica was measured to be an order of magnitude smaller than for the carbon sample. 

However, the measured charge bias characteristics are the only experimental data that provides a reference point for the simulations of the experiment dynamics. Since the spatial distribution of EUV intensity is subject to an unknown systematic measurement error, and the SEY is known to vary widely, depending on the surface conditions, we chose to keep the product of incident EUV  and SEY a constant, chosen to provide 1~nC total SE charge from the sample.

\subsection{Role of dielectric ring}
During the simulations we found that the mica accumulated charge over many EUV pulses, significantly effecting the local field distribution, and, hence changing the flux incident on the sample. If charging is neglected,  the Q~--~V response under vacuum and 2.8~Pa conditions cannot be reproduced for any reasonable parameter values. The discrepancy is caused by the potential barrier near the sample surface, which is created by the negative space charge generated during the EUV pulse. Although the potential is low, it produces a significant effect because the SE energy spectrum (\ref{eqn:se_spectrum}) is strongly peaked at rather low energies: approximately~\mbox{$W/3=1.3$~eV}.

\begin{figure}
    \centering
    \includegraphics[width=0.5\textwidth]{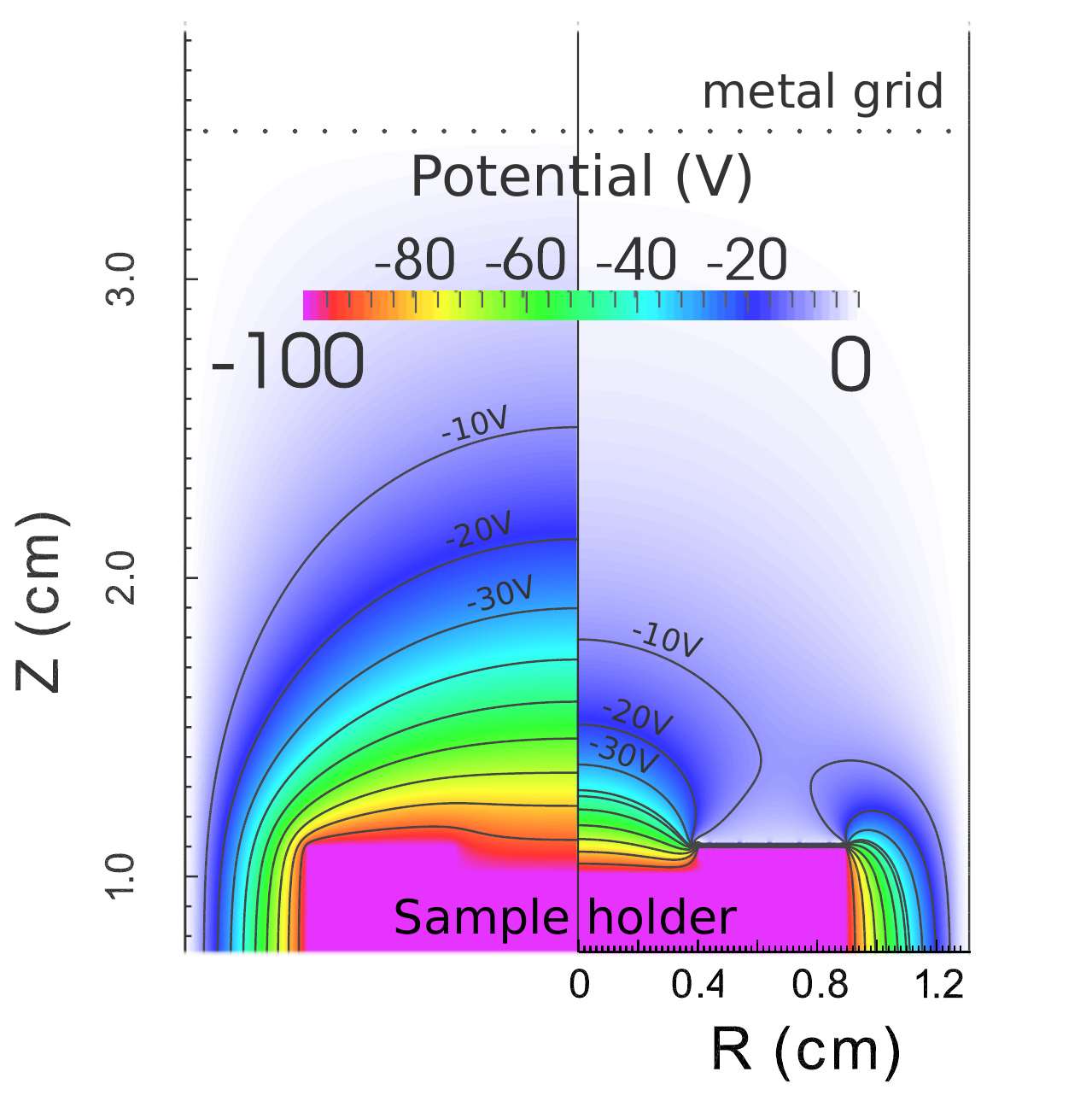}
    \caption{ Comparison of the potential distibution hear the sample holder for uncharged mica (left) and pre-charged mica for -100~V bias on the sample holder. The charge on the mica significantly increases the local field strength near the smaple surface. \label{fig:potential_no_plasma}}
\end{figure}

The  charge accumulation on the mica significantly increases the local field strength near the sample surface (see Fig.~\ref{fig:potential_no_plasma}). This removes the space charge potential barrier for biases of -100V bias and higher under the  conditions that we consider.   

The charging of the mica should saturate, leading to an unchanging charge density distribution on the mica. To estimate the charge density, we simulated one hundred pulses under  vacuum conditions for all experimentally applied  bias values. In these calculations,  the effective SEY from the mica is assumed to be about 0.001, which corresponds to the experimentally measured value.

This approach corresponds to the experimental procedure, since the Q~--~V characteristics were measured via averaging a large number of pulses for every combination of bias and pressure.  No special means were used in the experiment to  remove the accumulated charge from the mica between pulses.

For negative biases, simulations show that the mica potential rapidly reaches approximately 0~V potential. For these conditions, further charging is very slow, because (a) the electric field strength near the mica is very small, therefore, even a small amount of emitted charge creates a space charge potential barrier, and, (b), because the mica potential is close to zero, there is a small flux of electrons from the sample to the mica.

The continued slow variation in charge distribution is impractical to simulate, however.The final charge distribution requires a very long time to calculate and is  sensitive to the  combination of the SEY from both the sample and the mica, and to the corresponding  SE energy distribution functions. Nevertheless, the additional errors due to these limitations are expected to be small. In practice the maximum potential of the mica under  vacuum conditions  is bounded, since, as the mica's potential increases to $\sim$10~V, the electron current from the sample to mica becomes significant.

It is worth noting that the EUV plasma itself can contribute to charging  the mica, since, for  stationary discharges, the dielectric charges to the plasma potential. Despite the the fact that the potential of the EUV-induced plasma can be high (e.g. 20~--~30~V or more after EUV pulse), it rapidly decreases to several volts due to electron cooling due to collisions with the background gas. Most probably, the mica potential would converge to the time averaged potential of the plasma, e.g. several volts. Therefore, in the following results, the mica was pre-charged to the value that was obtained from simulations under vacuum conditions for the appropriate bias.

The optimum parameter combination that reproduces the experimentally measured Q~--~V characteristics were chosen as follows: the mica was pre-charged as described above, the EUV intensity was kept constant at the measured value, but the amount of the scattered EUV radiation incident on the sample holder was chosen so that 40$\%$ of EUV radiation was incident on the dielectric mica, and the effective SEY from the sample was kept at to 0.01 for the photon energy range of 60~eV~--~100~eV,  while the SE contribution from the VUV part of the spectrum was neglected. The simulated Q~--~V characteristic, after parameter optimization, and comparison with the measurements are presented in Fig.~\ref{fig:IV}.

\section{Ion fluxes to the sample surface}
The removal of carbon from the surface material should be directly dependent on the energy distribution function (EDF) of the ion flux incident on the surface. The conditions of the experimental study of carbon cleaning in the ISAN EUV experiment are summarized in Table~\ref{table:cleaning_conditions}. We computed the plasma conditions, ion fluxes, and ion EDF for these conditions. The simulated time for all cases was 10~$\mu$s. For 2.8~Pa~--~11.2~Pa and biases -200~V -- -100~V, the plasma had completely decayed and all ions were collected during the simulation time. But, for the 60~Pa and 86.5~Pa cases some plasma was still left in the simulation domain. 

\begin{figure}
    \centering
    \includegraphics[width=0.5\textwidth]{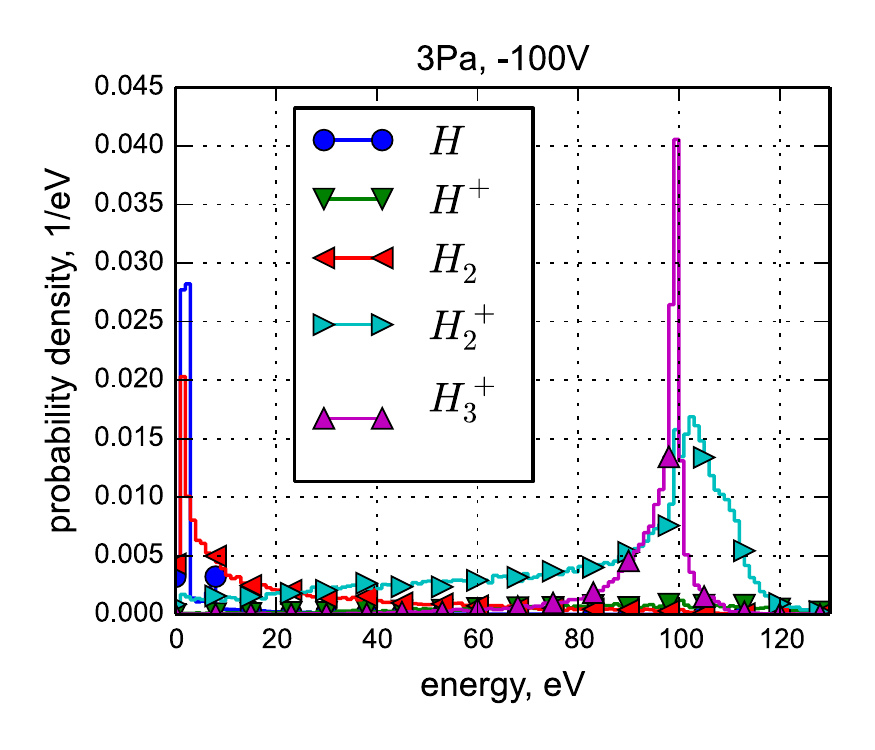} 
    \includegraphics[width=0.5\textwidth]{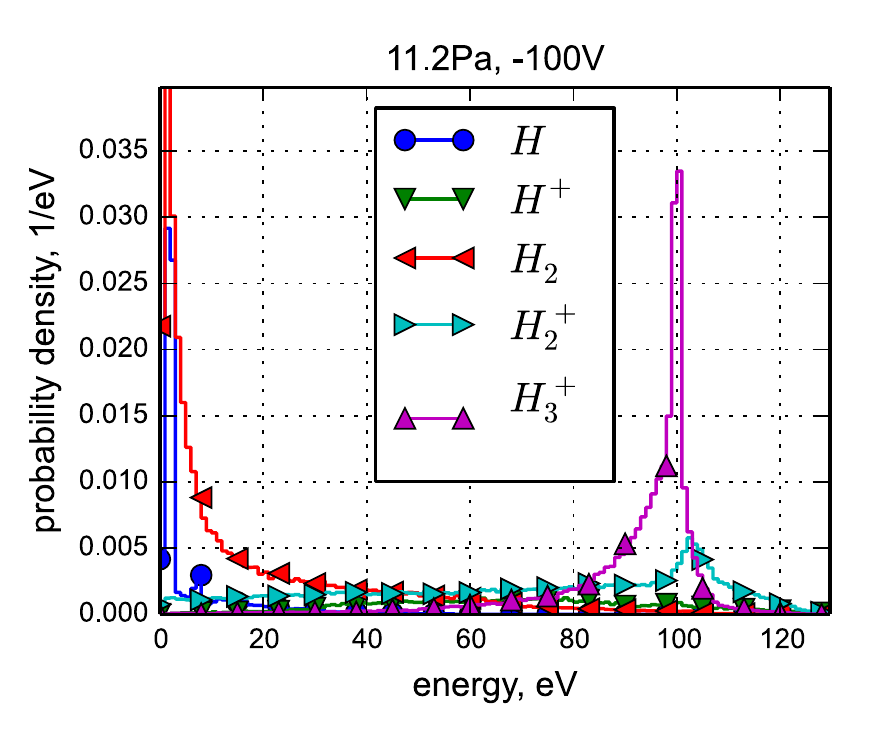}
    \caption{    
    Energy distribution function of ion and fast neutrals flux to the surface for 3~Pa (top) and 11.2~Pa (bottom) for -100V bias. The fluxes were integrated over time and over sample surface and sum of fluxes normalized to unity. Note that H$_3^+$ spike is on the same position over energy in spite of the increase of pressure. \label{fig:edf}}
\end{figure}

The EDFs, integrated over the simulation window,  and  averaged over the sample surface, are presented in Fig.~\ref{fig:edf}. The composition of the computed ion flux depends significantly on the pressure. For the 3~Pa case, the main ion is H$_2^+$, because the characteristic time for H$_2^+$ to H$_3^+$ conversion in 3~Pa H$_2$ is approximately 0.5~$\mu$s. With increasing pressure,  H$_3^+$ becomes the main ion, as expected. There is also a non-negligible contribution from fast H$_2$ which is produced due to a resonant charge exchange reaction. 

The maximum ion energy is larger than the applied bias voltage due to the build-up of the plasma potential. But the number of such fast ions is small, due to the fast plasma decay. For the same reason, the most energetic ion is H$^+$ because its small mass allows it to accelerate during the decay of the plasma potential.

It was observed, that contrary to the ion dose, the shape of the ion flux EDF was mainly defined by bias, pressure and the accumulated charge  on the dielectric. 
However, as discussed in previous section, we expect that the equilibrium mica potential does not significantly differ from 0~V. For small variations of the mica potential (if all other parameters are the same), the variations of the EDF shape are small for all simulations where a sample bias was applied. Moreover, the EDF was  barely sensitive to reasonable variations of other parameters.

This insensitivity is a consequence of the experimental procedure. The applied bias allows ions to be collected  from the entire chamber volume. Hence, the characteristics of the plasma are important only for the short  time when the sheath between plasma and sample is small. As ions are collected at the sample, the sheath size becomes larger and larger. Therefore, as time progresses, the instantantaneous ion flux EDF starts to depend only on the bias potential, mica potential, and the distance that the ions travel between the plasma and the sample. This is because these parameters determine the maximum possible energy of the ions, while the background pressure determines the number of collisions (e.g energy loss), and, hence, the average energy of the ion at the sample. Therefore, for the considered conditions, the shape of the ion flux EDF is mostly determined by the bias, mica potential, pressure, and the chamber geometry.

\begin{figure}
    \centering
    \includegraphics[width=0.5\textwidth]{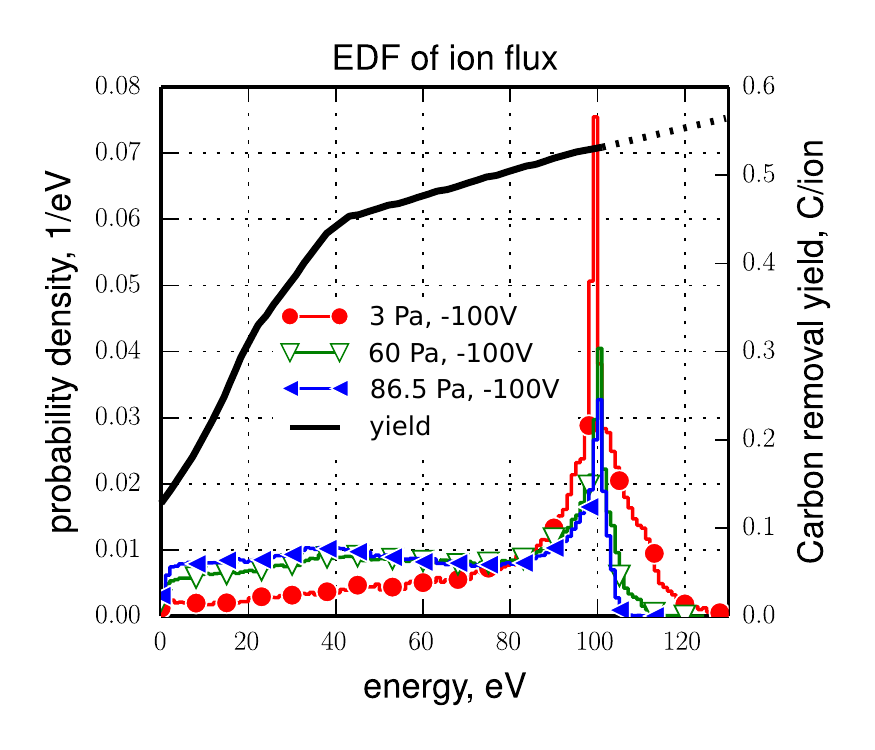}
    \caption{    
    Comparison of the  shape of the ion flux EDF obtained with pre-charged dielectric mica for different pressures. The ion fluxes were integrated over time and over sample surface, summed over ion types and normalized to unity. The carbon removal yield curve corresponds to surface wave discharge plasma (SWD) from~\cite{Dolgov.2013.comparison}. The dotted part of the yield curve corresponds to linear extrapolation.
    \label{fig:edf_comparison}} 
\end{figure}

\begin{figure}
    \centering
    \includegraphics[width=0.5\textwidth]{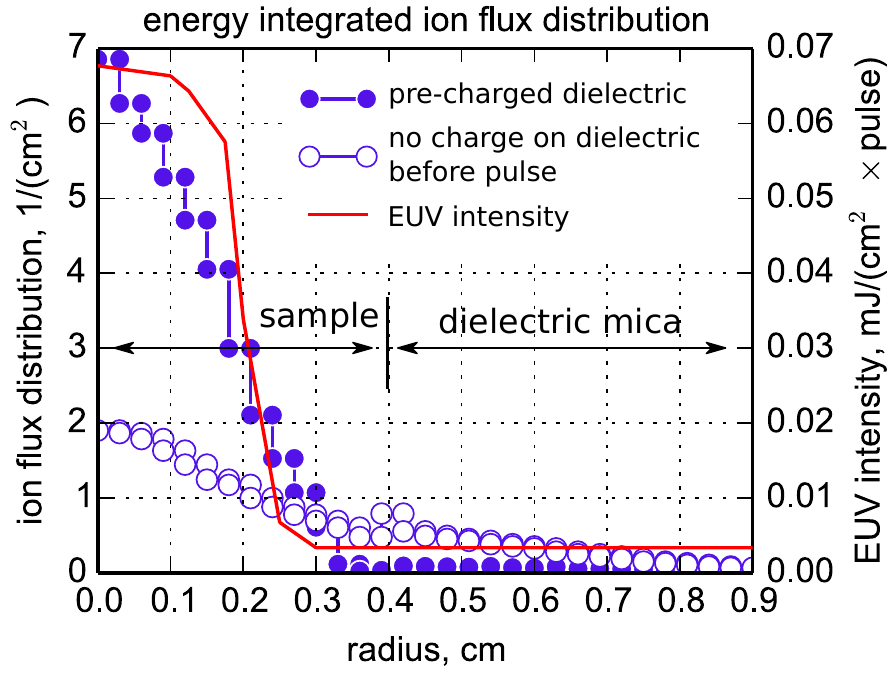}
    \caption{Comparison of the ion flux spatial distribution obtained with pre-charged mica and with zero charge on the mica. The ion fluxes were integrated over time and energy. Both distribution are normed to one if integrated over sample and dielectric mica surface areas. The conditions are 3~Pa background pressure and -100~V bias on the sample. The charge  on the mica focuses the ion flux on  the sample. Other combinations of pressure and bias show similar behavior. \label{fig:ion_flux_focusing} }
\end{figure}

It is notable, that the mica charge, due to exposure to EUV radiation and plasma, leads to the ion flux being focused on the sample (see Fig.~\ref{fig:ion_flux_focusing}). The ion flux to the mica itself becomes negligible  compared to the simulation with zero charge on the mica. Therefore, the measured ion doses should represent accurately the ion doses on the sample.

Hence, for the analysis of the experiments, it is reasonable to take the computed  EDF shape,  normed to the experimentally measured ion dose (if it is  available). This approach allows us to compare the experimental results at different pressures.

\section{Discussion \label{section:surface_model}} 
\begin{figure}  
  \centering
  \includegraphics[width=\columnwidth]{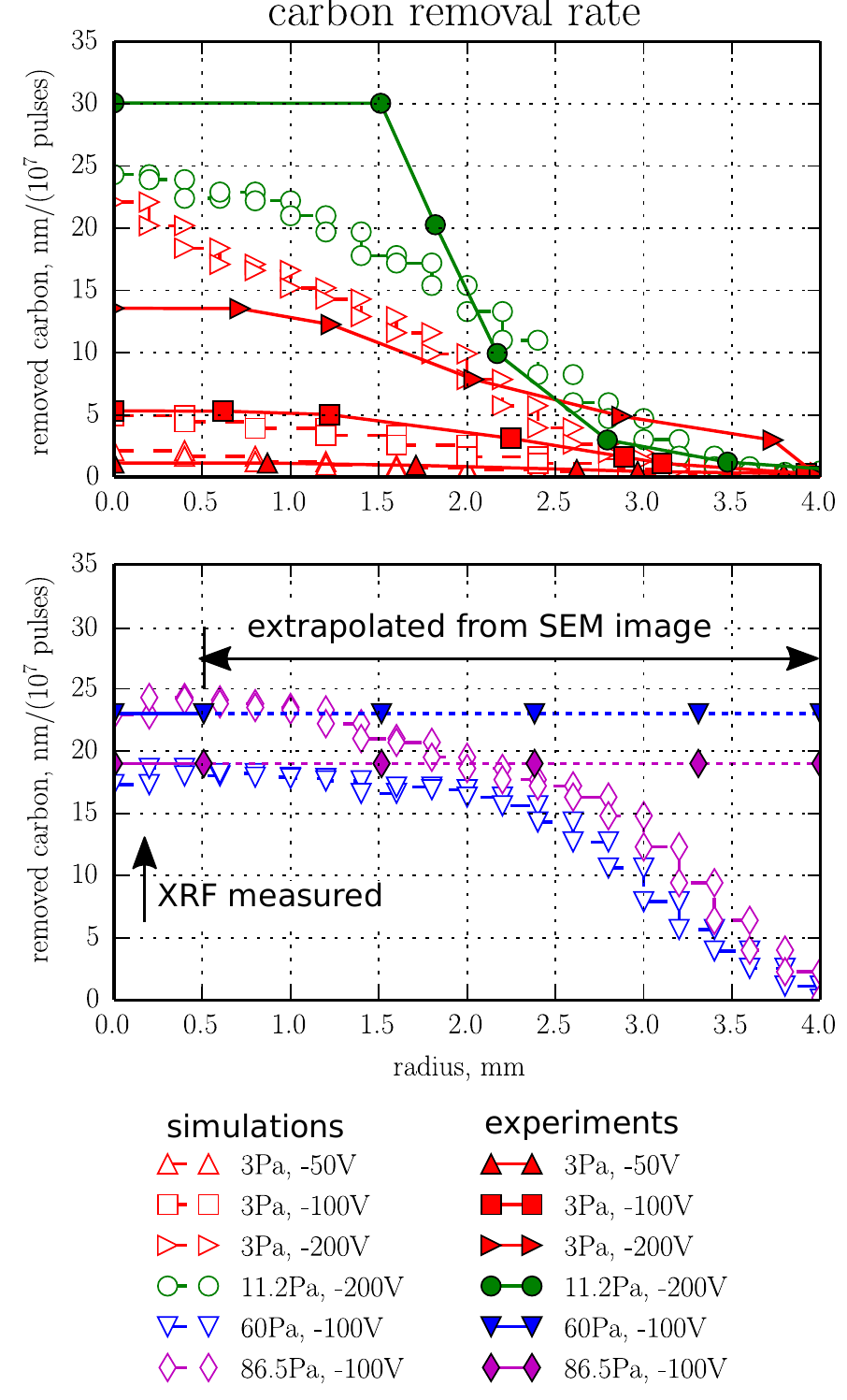}
  \caption{Experimentally observed and simulated carbon removal as function of pressure, bias and radius. For the conditions of 11.2~Pa, -200V the carbon layer was completely removed, exposing the substrate, at the center of the EUV spot. The etch yield for this case was set to the 1.6~C~atom/ion as in table \ref{table:cleaning_conditions}. For the conditions of 60~Pa, -200V, the carbon layer was completely removed from entire area exposed to plasma, hence the case is not shown. The simulated profiles for -50~V and -100~V bias were computed via convolution of the simulated ions fluxes with the yield from SWD experiments Fig.~\ref{fig:edf_comparison}. The profiles were normed to the ion dose per pulse from table~\ref{table:cleaning_conditions}.  The EUV dose was $10^7$ pulses per sample  in all cases, which is approximately 170~J. The EUV radiation was concentrated in a spot with radius approximately 2~mm. The detailed intensity distribution of the EUV spot is not accurately known.
\label{fig:proto_cleaning}}
\end{figure}

The  maximum etched depths and total etched volumes per exposure of 10$^7$ pulses for different pressures and biases are presented in Table.~\ref{table:cleaning_conditions}. The etched volumes are the integrals over etch profiles, which were determined by XRF measurement (see Fig.~\ref{fig:proto_cleaning}).
In the cases of 60~Pa and 86.5~Pa pressure, the etch profile was only measured at  the center of the EUV spot. We assumed, for these two cases, a uniform etching profile to estimate the carbon removal yield. Hence, for these cases, the amount of removed carbon is most probably overestimated.

\begin{table*}
  \begin{ruledtabular}
  \caption{Experimental conditions and carbon  removal rate \label{table:cleaning_conditions}}    
  \begin{tabular}{l l c c c c c c}
            &     &           & \multicolumn{2}{c}{removed carbon}              &   ion dose$^{b)}$    & \multicolumn{2}{c}{average yield}\\ 
pressure    & bias& EUV power &  max etched depth    & etched volume                &                  & present study &  SWD recomputed  \\
 Pa         &  V  &   W/cm$^2$& nm/$10^7$ pulse      & $10^{-6}$ cm$^3$/$10^7$pulse & nC/pulse         & C atom/ion    &  C atom/ion      \\
\hline                                                                                                                  
  3.0       & -50 &   0.11    & $1.7\pm0.5$          & $\sim0.03$                   & $\sim0.1$        &  0.5          &  0.4            \\
  3.0       & -100&   0.11    & $5.7\pm0.5$          & $\sim0.1$                    & $\sim0.25$       &  0.6          &  0.5           \\
  3.0       & -200&   0.13    & $14\pm0.5$           & $\sim0.3$                    & $\sim0.25$       &  1.9          &  ---            \\
  11.2      & -200&   0.13    & $30^{\,a)}\pm0.5$    & $\sim0.5$                    & $\sim0.5$        &  $>$1.6       &  ---            \\
    60      & -100&   0.13    & $23\pm0.5$           & $\sim 1.2^{\,c)}$            & $>2.2^{\,d)}$    &  $<$0.9       &  0.5            \\   
    60      & -200&   0.1     & $27\pm0.5^{\,e)}$    & $\sim 1.4^{\,c)}$            & $>2.3^{\,d)}$    &  $\sim$1.0${^{\,e)}}$&  ---     \\   
    86.5    & -100&   0.2     & $19\pm0.5$           & $\sim 1^{\,c)}$              & $>3.0^{\,d)}$    &  $<$0.5       &  0.4            \\          
  \end{tabular}
  \end{ruledtabular}
  \begin{flushleft}
  ${a)}$ The substrate was exposed, thus only lower boundary for cleaning rate could be estimated. \\     
  ${b)}$ Estimated from measured Q~--~V characteristics (see Fig.~\ref{fig:IV}) as the difference between the collected charge for a given pressure and bias voltage, and the collected charge for vacuum conditions.\\
  $c)$  The carbon removal was measured only in the center of the exposed area. However, an SEM image of the EUV exposed area showed no features. Therefore, for the purposes of estimating the carbon removal rate, we assume that the carbon was etched uniformly over the exposed area of the sample. \\
  ${d)}$ Experimental Q~--~V values were missing, therefore the simulated ion dose were used for estimation. Only ions which hit the sample during the simulated period are included. No corrections were made for the plasma which was present in the chamber after the end of the simulation.     \\ 
  $e)$ This value is only an order of magnitude  estimation, because  all carbon was removed from the sample.
  \end{flushleft}  
\end{table*}

As was discussed in the previous section, we used the experimentally measured ion doses  for yield estimation. However, for some experiments, the ion dose was not measured. In these cases, we used the ion doses obtained from simulation results. This approach is valid at high pressures because the uncertainty in the simulations is on the order of 30\%  (see Fig.~\ref{fig:IV}).

For all samples, the average carbon removal yield was larger than 0.2 carbon atoms per ion. Thus, physical sputtering is not the main process, since for energies below 200~eV, the expected sputtering yield is lower than \mbox{$5\cdot10^{-2}$} carbon atoms per hydrogen ion~\cite{YAMAMURA.1996.energy}. Therefore, carbon is removed via a chemical sputtering process, or a reactive ion etching process, because the effective yield for these processes is known to be large under certain conditions~\cite{Hopf.2003.chemical}. 

However, the results for -200~V bias, especially at low pressure (2.8~--~11.2~Pa), cannot be described by a chemical sputtering process, since the average yield is larger than one carbon atom per ion. Hence,  reactive ion etching may be  responsible for these very high etch rates. However, for  reactive ion etching there should be weakly bound radicals on the top of the carbon, which are desorbed from surface due to ion impact, which, in turn, lead to a very high yield per ion. Although, we have no data about these radicals, it is likely to be some form of methane radical (e.g., CH$_x$). The alternative: oxidation (e.g., CO and CO$_2$)~\cite{Hollenshead.2006.modelinga} is unlikely because water, which is the main source of oxygen, is removed from the hydrogen flow, and, during experiments, the chamber walls are kept at liquid nitrogen temperatures, trapping the majority of the residual water. Hence, the contribution of water to the carbon removal was limited by the initial coverage which is typically sub mono layer for amorphous carbon for the considered temperature range \cite{Gao.2014.defect}.

In the case of  experiments with  -50~V and -100~V biases, the results can be compared with  carbon etching in a surface wave discharge plasma~\cite{Dolgov.2013.comparison}, where similar magnetron deposited carbon on silicon substrate samples were used. The carbon removal yield for experiments with biases of -50~V and -100~V was estimated by convolving the simulated EDF with the energy dependent yield, as measured in the SWD experiments (see Fig.~\ref{fig:edf_comparison} and column "SWD recomputed" in table~\ref{table:cleaning_conditions}). In spite of  H$_3^+$ being the main ion in SWD plasma, we used the same yield for estimation, since we do not expect dramatic difference between H$_3^+$ and H$_2^+$ for carbon removal. These estimates show that the yields found in EUV experiments with low biases agree  with the recomputed yield  from the SWD experiment. The margin of error is, however, large enough that we cannot completely exclude direct influence from EUV radiation.

\section{Conclusion}
The evolution of an EUV induced hydrogen plasma was simulated. The simulations, due to their close coupling to experimental conditions, allowed the magnitude, composition, and energy spectrum of the flux from the plasma to the surface to be estimated. Our model successfully computes the charge bias characteristics of the sample to an accuracy of a factor of two.  This was used to quantitatively compare carbon etch rates between different EUV plasma conditions, as well as compare carbon etching under EUV-induced plasma to etching under SWD plasmas. In addition, the model describes the focusing effect of the dielectric surround, which allows the measured carbon etch profiles to be better understood in a quantitative manner.

It was observed that the carbon etching mechanism at low bias and pressure was different than that at high bias and pressure. For the higher energy range, the carbon removal yield in EUV-induced plasma was larger than one carbon atom per ion. Most probably, etching is dominated by a reactive ion etching process, which may be due to the production of methane radicals (e.g. CH$_x$)  that can desorb under energetic ion flux.

However, at low  bias energies, the etch rates found in EUV-induced plasma agree with those found in SWD plasma to within model and measurement uncertainties. Therefore, the role of the EUV radiation in a previous study~\cite{Dolgov.2013.comparison} was overestimated at low bias energies and pressures.

However, our study does not exclude EUV enhancing etch rates at low biases or having a small ($< 0.2$~C/ion) enhancing effect over the range of biases and pressures that we studied. Therefore,  experiments to determine the effect of EUV  in the limit of low-to-no bias are required.

\begin{acknowledgements}
The authors would like to thank Tatiana Rakhimova (Skobeltsyn Institute of Nuclear Physics) for helpful discussion.  

This work is part of the research program ``Controlling photon and plasma induced processes at EUV optical surfaces (CP3E)'' of the ``Stichting voor Fundamenteel Onderzoek der Materie (FOM)'' which is financially supported by the Nederlandse Organisatie voor Wetenschappelijk Onderzoek (NWO).
The CP3E programme is cofinanced by industrial partners, including ASML (Veldhoven), and the AgentschapNL through the Catrene EXEPT program. 
\end{acknowledgements}

\bibliographystyle{unsrtnat}
\bibliography{refs}

\begin{thebibliography}{31}
\providecommand{\natexlab}[1]{#1}
\providecommand{\url}[1]{\texttt{#1}}
\expandafter\ifx\csname urlstyle\endcsname\relax
  \providecommand{\doi}[1]{doi: #1}\else
  \providecommand{\doi}{doi: \begingroup \urlstyle{rm}\Url}\fi

\bibitem[Boller et~al.(1983)Boller, Haelbich, Hogrefe, Jark, and
  Kunz]{Boller.1983.investigation}
K.~Boller, R.~P. Haelbich, H.~Hogrefe, W.~Jark, and C.~Kunz.
\newblock Investigation of carbon contamination of mirror surfaces exposed to
  synchrotron radiation.
\newblock \emph{Nuclear Instruments and Methods in Physics Research},
  208\penalty0 (1{\textendash}3):\penalty0 273--279, 1983.
\newblock ISSN 0167-5087.
\newblock \doi{10.1016/0167-5087(83)91134-1}.

\bibitem[Hollenshead and Klebanoff(2006)]{Hollenshead.2006.modelinga}
Jeromy Hollenshead and Leonard Klebanoff.
\newblock Modeling extreme ultraviolet/{H2O} oxidation of ruthenium optic
  coatings.
\newblock \emph{Journal of Vacuum Science \& Technology B: Microelectronics and
  Nanometer Structures}, 24\penalty0 (1):\penalty0 118--130, 2006.
\newblock \doi{10.1116/1.2150225}.

\bibitem[Nakayama et~al.(2009)Nakayama, Miyake, Takase, Terashima, Sudo,
  Watanabe, and Fukuda]{Nakayama.2009.analysis}
Takahiro Nakayama, Akira Miyake, Hiromitsu Takase, Shigeru Terashima, Takashi
  Sudo, Yutaka Watanabe, and Yasuaki Fukuda.
\newblock Analysis of carbon deposition on multilayer mirrors by using two
  different beamlines.
\newblock In Frank~M. Schellenberg and Bruno~M. La~Fontaine, editors,
  \emph{Alternative Lithographic Technologies}, volume 7271, pages 72713P--8,
  San Jose, {CA}, {USA}, 2009. {SPIE}.

\bibitem[Shin et~al.(2009)Shin, Sporre, Raju, and
  Ruzic]{Shin.2009.reflectivity}
H~Shin, J~Sporre, R~Raju, and D~Ruzic.
\newblock Reflectivity degradation of grazing-incident {EUV} mirrors by {EUV}
  exposure and carbon contamination.
\newblock \emph{Microelectronic Engineering}, 86\penalty0 (1):\penalty0
  99--105, January 2009.
\newblock ISSN 01679317.
\newblock \doi{10.1016/j.mee.2008.10.009}.

\bibitem[Davis et~al.(2007)Davis, Kyriakou, Grant, Tikhov, and
  Lambert]{Davis.2007.situ}
David~J. Davis, Georgios Kyriakou, Robert~B. Grant, Mintcho~S. Tikhov, and
  Richard~M. Lambert.
\newblock Toward the in situ remediation of carbon deposition on ru-capped
  multilayer mirrors intended for {EUV} lithography:\hspace{0.167em} exploiting
  the electron-induced chemistry.
\newblock \emph{The Journal of Physical Chemistry C}, 111\penalty0
  (33):\penalty0 12165--12168, 2007.
\newblock \doi{10.1021/jp074766y}.

\bibitem[Hopf et~al.(2003)Hopf, von Keudell, and Jacob]{Hopf.2003.chemical}
C.~Hopf, A.~von Keudell, and W.~Jacob.
\newblock Chemical sputtering of hydrocarbon films.
\newblock \emph{Journal of Applied Physics}, 94\penalty0 (4):\penalty0
  2373--2380, August 2003.
\newblock ISSN 00218979.
\newblock \doi{doi:10.1063/1.1594273}.

\bibitem[K{\"u}ppers(1995)]{Kuppers.1995.hydrogen}
J{\"u}rgen K{\"u}ppers.
\newblock The hydrogen surface chemistry of carbon as a plasma facing material.
\newblock \emph{Surface Science Reports}, 22\penalty0 (7-8):\penalty0 249--321,
  1995.
\newblock ISSN 0167-5729.
\newblock \doi{10.1016/0167-5729(96)80002-1}.

\bibitem[Liu et~al.(2010)Liu, Sun, Dai, Stirner, and Wang]{Liu.2010.general}
Shengguang Liu, Jizhong Sun, Shuyu Dai, Thomas Stirner, and Dezhen Wang.
\newblock A general model for chemical erosion of carbon materials due to
  low-energy h+ impact.
\newblock \emph{Journal of Applied Physics}, 108\penalty0 (7):\penalty0 073302,
  2010.
\newblock ISSN 00218979.
\newblock \doi{10.1063/1.3485821}.

\bibitem[Jariwala et~al.(2009)Jariwala, Ciobanu, and
  Agarwal]{Jariwala.2009.atomic}
Bhavin~N. Jariwala, Cristian~V. Ciobanu, and Sumit Agarwal.
\newblock Atomic hydrogen interactions with amorphous carbon thin films.
\newblock \emph{Journal of Applied Physics}, 106\penalty0 (7):\penalty0 073305,
  2009.
\newblock ISSN 00218979.
\newblock \doi{10.1063/1.3238305}.

\bibitem[Dolgov et~al.(2014)Dolgov, Lopaev, Rachimova, Kovalev, Vasilyeva, Lee,
  Krivtsun, Yakushev, and Bijkerk]{Dolgov.2013.comparison}
A.~Dolgov, D.~Lopaev, T.~Rachimova, A.~Kovalev, A.~Vasilyeva, C.~J. Lee, V.~M.
  Krivtsun, O.~Yakushev, and F.~Bijkerk.
\newblock Comparison of h2 and he carbon cleaning mechanisms in extreme
  ultraviolet induced and surface wave discharge plasmas.
\newblock \emph{Journal of Physics D: Applied Physics}, 47\penalty0
  (6):\penalty0 065205, 2014.
\newblock \doi{10.1088/0022-3727/47/6/065205}.

\bibitem[Dolgov et~al.(2015)Dolgov, Yakushev, Abrikosov, Snegirev, Krivtsun,
  Lee, and Bijkerk]{Dolgov.2015.extreme}
A~Dolgov, O~Yakushev, A~Abrikosov, E~Snegirev, V~M Krivtsun, C~J Lee, and
  F~Bijkerk.
\newblock Extreme ultraviolet ({EUV}) source and ultra-high vacuum chamber for
  studying {EUV}-induced processes.
\newblock \emph{Plasma Sources Science and Technology}, 24\penalty0
  (3):\penalty0 035003, June 2015.
\newblock ISSN 0963-0252, 1361-6595.
\newblock \doi{10.1088/0963-0252/24/3/035003}.

\bibitem[Bakshi(2006)]{Bakshi.2006.euv_sources}
Vivek Bakshi.
\newblock \emph{{EUV} sources for lithography}.
\newblock {SPIE} Press, Bellingham, Wash., 2006.
\newblock ISBN 978-1-61583-716-8 1-61583-716-7 978-0-8194-5845-2 0-8194-5845-7.

\bibitem[Birdsall and Langdon(1985)]{Birdsall.1985.plasma}
Charles~K Birdsall and A.~Bruce Langdon.
\newblock \emph{Plasma physics via computer simulation}.
\newblock {McGraw}-Hill, New York, 1985.
\newblock ISBN 0-07-005371-5 978-0-07-005371-7.

\bibitem[Dunaevsky et~al.(2003)Dunaevsky, Raitses, and
  Fisch]{Dunaevsky.2003.secondary}
A.~Dunaevsky, Y.~Raitses, and N.~J. Fisch.
\newblock Secondary electron emission from dielectric materials of a hall
  thruster with segmented electrodes.
\newblock \emph{Physics of Plasmas}, 10\penalty0 (6):\penalty0 2574, 2003.
\newblock ISSN 1070664X.
\newblock \doi{10.1063/1.1568344}.

\bibitem[Fowler and Farnsworth(1958)]{Fowler.1958.reflection}
H.~A. Fowler and H.~E. Farnsworth.
\newblock Reflection of very slow electrons.
\newblock \emph{Physical Review}, 111\penalty0 (1):\penalty0 103--112, 1958.
\newblock \doi{10.1103/PhysRev.111.103}.

\bibitem[van~der Velden et~al.(2006)van~der Velden, Brok, van~der Mullen,
  Goedheer, and Banine]{vanderVelden.2006.particle-in-cell}
M.~H.~L. van~der Velden, W.~J.~M. Brok, J.~J. A.~M. van~der Mullen, W.~J.
  Goedheer, and V.~Banine.
\newblock Particle-in-cell monte carlo simulations of an extreme ultraviolet
  radiation driven plasma.
\newblock \emph{Physical Review E}, 73\penalty0 (3):\penalty0 036406, March
  2006.
\newblock \doi{10.1103/PhysRevE.73.036406}.

\bibitem[Langmuir(1923)]{Langmuir.1923.effect}
Irving Langmuir.
\newblock The effect of space charge and initial velocities on the potential
  distribution and thermionic current between parallel plane electrodes.
\newblock \emph{Physical Review}, 21\penalty0 (4):\penalty0 419--435, April
  1923.
\newblock \doi{10.1103/PhysRev.21.419}.

\bibitem[Henke et~al.(1977)Henke, Smith, and
  Attwood]{Henke.1977.electron_emission}
Burton~L. Henke, Jerel~A. Smith, and David~T. Attwood.
\newblock 0.1{\textendash}10-{keV} x-ray-induced electron emissions from
  solids{\textemdash}models and secondary electron measurements.
\newblock \emph{Journal of Applied Physics}, 48:\penalty0 1852, 1977.
\newblock ISSN 00218979.
\newblock \doi{10.1063/1.323938}.

\bibitem[Lide(2003)]{Lide.2003.crc}
David~R Lide.
\newblock \emph{{CRC} handbook of chemistry and physics, 2003-2004: a
  ready-reference book of chemical and physical data.}
\newblock {CRC} Press, Boca Raton, Fla., 2003.
\newblock ISBN 0-8493-0484-9 978-0-8493-0484-2.

\bibitem[Chung et~al.(1993)Chung, Lee, Masuoka, and
  Samson]{Chung.1993.dissociative}
Y.~M Chung, E.~M Lee, T.~Masuoka, and James A.~R Samson.
\newblock Dissociative photoionization of h2 from 18 to 124 {eV}.
\newblock \emph{The Journal of Chemical Physics}, 99\penalty0 (2):\penalty0
  885--889, July 1993.
\newblock ISSN 00219606.
\newblock \doi{doi:10.1063/1.465352}.

\bibitem[Mokrov and Raizer(2008)]{Mokrov.2008.monte_carlo}
M.~S. Mokrov and Yu~P. Raizer.
\newblock Monte carlo method for finding the ionization and secondary emission
  coefficients and i{\textendash}v characteristic of a townsend discharge in
  hydrogen.
\newblock \emph{Technical Physics}, 53\penalty0 (4):\penalty0 436--444, April
  2008.
\newblock ISSN 1063-7842, 1090-6525.
\newblock \doi{10.1134/S1063784208040075}.

\bibitem[Brunger and Buckman(2002)]{Brunger.2002.electronmolecule}
M.~J. Brunger and S.~J. Buckman.
\newblock Electron{\textendash}molecule scattering cross-sections. i.
  experimental techniques and data for diatomic molecules.
\newblock \emph{Physics reports}, 357\penalty0 (3):\penalty0 215--458, 2002.

\bibitem[Shyn et~al.(1981)Shyn, Sharp, and Kim]{Shyn.1981.doubly}
T.~W. Shyn, W.~E. Sharp, and Y.-K. Kim.
\newblock Doubly differential cross sections of secondary electrons ejected
  from gases by electron impact: 25-250 {eV} on h2.
\newblock \emph{Physical Review A}, 24\penalty0 (1):\penalty0 79--88, July
  1981.
\newblock \doi{10.1103/PhysRevA.24.79}.

\bibitem[Rudd et~al.(1993)Rudd, Hollman, Lewis, Johnson, Porter, and
  Fagerquist]{Rudd.1993.doubly}
M.~E. Rudd, K.~W. Hollman, J.~K. Lewis, D.~L. Johnson, R.~R. Porter, and E.~L.
  Fagerquist.
\newblock Doubly differential electron-production cross sections for
  200{\textendash}1500-{eV} {e+H2} collisions.
\newblock \emph{Physical Review A}, 47\penalty0 (3):\penalty0 1866--1873, March
  1993.
\newblock \doi{10.1103/PhysRevA.47.1866}.

\bibitem[{\v S}imko et~al.(1997){\v S}imko, Marti{\v s}ovit{\v s}, Bretagne,
  and Gousset]{Simko.1997.transport}
T.~{\v S}imko, V.~Marti{\v s}ovit{\v s}, J.~Bretagne, and G.~Gousset.
\newblock Computer simulations of h+ and h3+ transport parameters in hydrogen
  drift tubes.
\newblock \emph{Physical Review E}, 56\penalty0 (5):\penalty0 5908--5919,
  November 1997.
\newblock \doi{10.1103/PhysRevE.56.5908}.

\bibitem[Nanbu(1994)]{Nanbu.1994.simple}
Kenichi Nanbu.
\newblock Simple method to determine collisional event in monte carlo
  simulation of electron-molecule collision.
\newblock \emph{Japanese Journal of Applied Physics}, 33\penalty0 (Part 1, No.
  8):\penalty0 4752--4753, August 1994.
\newblock ISSN 0021-4922.
\newblock \doi{10.1143/JJAP.33.4752}.

\bibitem[Dutton(1975)]{Dutton.1975.survey}
J.~Dutton.
\newblock A survey of electron swarm data.
\newblock \emph{Journal of Physical and Chemical Reference Data}, 4\penalty0
  (3):\penalty0 577--856, July 1975.
\newblock ISSN 0047-2689, 1529-7845.
\newblock \doi{10.1063/1.555525}.

\bibitem[Graham et~al.(1973)Graham, James, Keever, Albritton, and
  {McDaniel}]{Graham.1973.mobilities}
E.~Graham, D.~R. James, W.~C. Keever, D.~L. Albritton, and E.~W. {McDaniel}.
\newblock Mobilities and longitudinal diffusion coefficients of mass-identified
  hydrogen ions in h2 and deuterium ions in d2 gas.
\newblock \emph{The Journal of Chemical Physics}, 59\penalty0 (7):\penalty0
  3477--3481, October 1973.
\newblock ISSN 00219606.
\newblock \doi{doi:10.1063/1.1680505}.

\bibitem[Yakshinskiy et~al.(2007)Yakshinskiy, Wasielewski, Loginova, and
  Madey]{Yakshinskiy.2007.carbon}
B.~V. Yakshinskiy, R.~Wasielewski, E.~Loginova, and Theodore~E. Madey.
\newblock Carbon accumulation and mitigation processes, and secondary electron
  yields of ruthenium surfaces.
\newblock In \emph{Proceedings of {SPIE}}, pages 65172Z--65172Z--11, San Jose,
  {CA}, {USA}, 2007.
\newblock \doi{10.1117/12.711785}.

\bibitem[{YAMAMURA} and {TAWARA}(1996)]{YAMAMURA.1996.energy}
{YASUNORI} {YAMAMURA} and {HIRO} {TAWARA}.
\newblock {ENERGY} {DEPENDENCE} {OF} {ION}-{INDUCED} {SPUTTERING} {YIELDS}
  {FROM} {MONATOMIC} {SOLIDS} {AT} {NORMAL} {INCIDENCE}.
\newblock \emph{Atomic Data and Nuclear Data Tables}, 62\penalty0 (2):\penalty0
  149--253, March 1996.
\newblock ISSN 0092-640X.
\newblock \doi{10.1006/adnd.1996.0005}.

\bibitem[Gao et~al.(2014)Gao, Zoethout, Sturm, Lee, and
  Bijkerk]{Gao.2014.defect}
A.~Gao, E.~Zoethout, J.M. Sturm, C.J. Lee, and F.~Bijkerk.
\newblock Defect formation in single layer graphene under extreme ultraviolet
  irradiation.
\newblock \emph{Applied Surface Science}, 317:\penalty0 745--751, October 2014.
\newblock ISSN 01694332.
\newblock \doi{10.1016/j.apsusc.2014.08.177}.

\end{thebibliography}
\end{document}